\documentclass[12pt]{article}
\usepackage{bbold}
\usepackage{graphicx}
\usepackage{dcolumn}
\usepackage{bm}
\usepackage{pstricks}
\usepackage{amsfonts}
\usepackage{amssymb}
\usepackage[T1]{fontenc}

\newcommand{\be}{\begin{equation}}
\newcommand{\ee}{\end{equation}}
\newcommand{\bea}{\begin{eqnarray}}
\newcommand{\eea}{\end{eqnarray}}

\newcommand{\zt}{\tau}
\newcommand{\zE}{I\hskip-3.7pt E}

\newcommand{\zN}{{\mathbb N}} 
\newcommand{\zR}{{\mathbb R}}
\newcommand*{\zII}{{\normalfont\hbox{1\kern-0.185em \vrule width .4pt depth .0pt}}}

\newcommand {\bdm} {\begin{displaymath}}
\newcommand {\edm} {\end{displaymath}}
\newcommand {\ba}  {\begin{array}}
\newcommand {\ea}  {\end{array}}

\begin{document}

\newtheorem{thm}{Theorem}
\newtheorem{cor}{Corollary}
\newtheorem{Def}{Definition}
\newtheorem{lem}{Lemma}
\begin{center}

{\large \bf Brownian Motion and General Relativity} \vspace{5mm}

Paul O'Hara$^1$ and Lamberto Rondoni$^{2,3}$
\\
\vspace{5mm}
{\small\it
$^1$Dept. of Mathematics, Northeastern Illinois University, 5500 North St. Louis Avenue, Chicago,
Illinois 60625-4699, USA \\
$^2$ Dip. Scienze Matematiche, Politecnico di Torino, C. Duca degli Abruzzi 24, 10129 Torino \\
$^3$ INFN, Sezione di Torino, Via P. Giuria 1, 10125 Torino, Italy}
\end{center}
\vspace{10mm}
\begin{abstract}
We construct a model of Brownian Motion on a pseudo-Riemannian
manifold associated with general relativity. There are two aspects of the problem: The first is to define
a sequence of stopping times associated with the Brownian ``kicks'' or impulses. The second is
to define the dynamics of the particle along geodesics in between the Brownian kicks. When these two
aspects are taken together, we can associate various distributions with the motion.
We will find that the statistics of space-time events will obey a temperature dependent
four dimensional Gaussian distribution defined over the quaternions which locally can be 
identified with Minkowski space.
Analogously, the statistics of the 4-velocities
will obey a kind of Maxwell-Juttner distribution.
In contrast to previous work, our processes are characterized by two independent proper time variables 
defined with respect to the laboratory frame: a discrete one corresponding to the stopping times when 
the impulses take place and a continuous one corresponding to the geodesic motion in-between impulses. 
The subsequent distributions are then solutions of partial differential equations which contain 
derivatives with respect to both time variables.
\end{abstract}

\noindent KEY WORDS: geodesic and non-geodesic motion, Gaussian and Maxwell-J\"uttner distributions, stopping times


\section{Introduction}
Brownian motion is one of the cornerstones of Statistical Mechanics. Einstein successfully
used it to give a rational proof of the existence of atoms \cite{AE05}, and since then
our understanding of it has been used
as a paradigm to model systems in contact with a heat reservoir. It has also been
used as a stochastic model to represent a variety of different phenomena in such diverse
fields as physics, chemistry, biology, finance etc. Indeed, its universal character rests on it being the simplest
model available for describing time evolution implied by a combination of random and deterministic
factors \cite{Dup,BPRV08,DH09}.
In the case of Einstein's theory, the deterministic factor is given by the Stokes force exerted on pollen
grains by a liquid, seen as a continuum macroscopic medium; while the random factor represents the
impulses given to the same pollen grains by the myriad of fluid molecules colliding with them.
Einstein's ingenuity consisted in understanding that water could be seen as acting on
pollen grains in two almost antithetical ways: as a continuum with its viscosity (systematic
component), and as a collection of many interacting particles (chance). Clearly, many phenomena can be
interpreted as the result of the cooperation of systematic and random events, hence the success of
his simple model in combining the two. For example, in the case of a tagged molecule of a gas, the free flights between collisions with
other molecules constitute the systematic part, while the chance collisions with other molecules which interrupt the free flights, constitute the non-systematic component of the motion causing it to move
in a random environment \cite{CCFV}.\footnote{Equivalent situations are realized with particles tracing 
deterministic trajectories in regular environments, if correlations decay in time and space, making inapplicable 
a deterministic description. This happens, for instance, in the so-called periodic Lorentz gas, consisting of 
point particles moving in a
periodic array of convex (typically circular) scatterers, in which position and velocity correlations
decay at an exponential rate, \cite{MR94,LNRM}. Another example is given by polygonal billiards, in which 
correlations do not decay exponentially fast \cite{JR06}. In that case, one observes a different
class of phenomena, which imply anomalous rather than standard diffusion.} In principle, this picture
applies to all forms of dynamics, including special and general relativity.

One could thus explore the possibility of addressing the relativistic Brownian motion, as a random
process with stationary independent increments, in which a Brownian particle travels on a geodesic
until it is dislodged by the interaction with another (moving or standing) object, which shifts it onto
a new geodesic.

Seen from this perspective, there are two aspects to consider. One is connected with the specific dynamics of the
particle along the smooth parts of its piecewise smooth trajectory, while the second is connected with the random
fluctuations that occur as the particle bounces from one smooth section of the trajectory to another.
In practice, the geodesic motion between the interactions obeys the deterministic laws of relativistic dynamics,
while the collection of impulses assigns random orientations, velocities and accelerations according to appropriate
statistical laws, although the actual accelerations along geodesics will be determined by the dynamics in between collisions.

This perspective will enable us to overcome the historical difficulties associated with defining a Brownian motion 
within the framework of relativity. For example, a straightforward extension of the classical theory to the 
relativistic one is known to
be frustrated in practice by the difficulty (perhaps impossibility) of expressing the dynamics of
interacting many body systems within the relativistic framework. The choice of reference frame becomes problematic 
especially
when accelerations are involved. Furthermore, it is hard to formulate the interaction potential because the action and 
reaction principle holds
only for contact interactions; and the effects of length contraction and time dilation add to the puzzle
\cite{DH09,ARM,GM11}. Also from the point of view of stochastic processes and of the diffusion processes associated 
with Brownian motion, with the exception of a non-trivial time-discrete relativistic Markov model found in
\cite{dth07}, certain relativistic generalizations and their Gaussian solutions must necessarily be non-Markovian 
or reduce to singular functions (cf.\ the excellent Review \cite{DH09}, and references therein). Indeed, this has been 
investigated in great detail by Hakim \cite{hakim68} who defines relativistic stochastic processes 
in $\mu = \mathcal{M}^4 \times U^4$ where ${\cal M}^4$ is the Minkowski space-time and $U^4$ is the space of 
velocity 4-vectors but shows it is not suitable for defining a Markov process.\footnote{Hakim manages to define 
relativistic stochastic processes by indexing the stochastic events with the set of all regular
7-dimensional space-like hypersurfaces \cite{hakim68}. However, that set is only partially ordered, 
hence not suitable for the Markov property.}
In addition, there is a theorem by Dvoretzky, Erd\"os and Kakutani which states that for Brownian motion the 
sample paths are almost surely nowhere differentiable and one may wonder how this fits into the overall 
framework of relativity and differential manifolds.

Conceptually, these kind of difficulties can be bypassed, if one is not
interested in a detailed description of the interaction processes \cite{CCRel,GGbook}
and treats the impulses
as a set of random variables occurring at random discrete proper times.\footnote{For instance, in the low density
limit, in which the interaction (potential) energy is negligible compared to the kinetic energy,
Ref.\cite{ARM} describes a relativistic gas as a collection of particles which move according to
special relativity from collision to collision, and treats as classical the ``randomly'' occurring 
collisions among particles.
In this way Ref.\cite{ARM} provides numerically a dynamical justification of the hypothesis of
molecular chaos underlying the validity of the relativistic Boltzmann equation and of its equilibrium
solution known as the Maxwell-J\"uttner distribution \cite{DGRel,CCRel}.}

However, to cast a deterministic process, such as the one described by Ref.\cite{ARM}, in the form
of a stochastic process, one cannot index the events by their proper time alone. Instead, it is necessary to 
distinguish the random parts from the deterministic parts of the motion by means of stopping times and then, 
in order to be coherent, refer all motions to the same (laboratory) frame of reference. Three cases arise. 
We can have:
\begin{itemize}
\item[1)] A Brownian random walk in which a single time scale is used and both the stopping times and the time 
between jumps are correlated by means of a common time. For example, Keller in \cite{Keller} considers a particle 
moving 
along the $x$ axis such that during a time interval of duration $\tau=1$ from $t=i-1$ to $t=i$, $i=1,2,\dots$, 
the particle moves with velocity $\nu =+1$ or $\nu =-1$, each with probability $\frac{1}{2}$.\newline
\item[2)] A Brownian random walk in which two independent time scales are used, with each stopping time indexed 
by the time of the Brownian kicks as measured in the rest frame of the Brownian particle. In other words,
the process obeys the strong Markov property and the proper time between consecutive stopping times is a random 
variable, with the Brownian flights lying along
a piecewise differentiable curve of random length each with its own proper time. 
\newline
\item[3)] A continuous time Brownian motion in which two independent proper time scales are used with stopping 
times being chosen such that the time between two consecutive stopping times is an arbitrary constant (not a 
random variable), but with distance and velocity between stopping times defining a stationary independent 
Markov process.
\end{itemize}
For the purpose of this article, we focus on case (2), leaving case (3) for a later date. We propose a new perspective, based on combining the discrete stopping times and the continuous time associated with the deterministic part of the motion. 
By using the proper time, the same event viewed from different frames can be synchronized. Indeed, from the perspective of the rest frame (heat bath) of the particle, Brownian impulses recorded at $\tau_1, \tau_2, \dots \tau_i \dots $ can be transmitted to the laboratory frame. For example, the numbers .5, .23, .678, .45, ...  could represent the times between succesive impulses as recorded in the rest frame, which for a large sample can be used in principle to calculate the variance of $\tau_i-\tau_{i-1}$. Since each $\tau_i-\tau_{i-1}$ is a proper time (recall the local time of the rest frame is equivalent to the proper time) it remains invariant in general relativity under coordinate transformations, which also includes the laboratory frame.  This is the only information transmitted from frame to frame. Moreover, if we assume that the heat bath is held at a constant temperature, then it is also reasonable to assume using the strong Markov property 
that the set $\{\tau_i-\tau_{i-1}\}_{i=1}^\infty$ 
defines a set of stationary independent increments. This statement is independent of reference frames.

It is also true that from the laboratory frame's perspective we do not know the local coordinates $(t,x,y,z)$ of the 
Brownian particle and that many options are possible.  However, we can extract enough information from the invariance 
of 
the $\{\tau_i-\tau_{i-1}\}$ and the Central Limit Theorem to specify the probability of finding the particle in any 
given 
region defined with respect to a reference frame. Our density function will be Lorentz invariant (see below) and 
consequently allows us to calculate probabilities in any reference frame. Indeed, that is why we are formulating 
the theory from the statistical perspective and not deterministic mechanics. 

To conclude, the novelty of this article will be to show that a relativistic Brownian motion can be described 
when motion between the discrete random stopping times lies along a geodesic path. In doing so, we obtain a 
covariant description of the relativistic Brownina motion, analogously to Ref.\cite{Larral}, and we 
also circumvent the difficulty in dealing with nowhere differentiable paths associated with case (3). 

For the model developed 
here, we show that in the case of general or special relativity, Brownian motion can exist along piecewise 
geodesic curves on a pseudo-Riemannian manifold.

\section{Working with geodesics}
The key to the above mentioned developments will be found in working with motion along geodesics. Beginning with the metric tensor
\begin{equation} ds^2=g_{ij}dx^idx^j \label{ds^2}\end{equation}
this can be re-written as
$$ds=g_{ij}\frac{dx^i}{ds}dx^j.$$
However, from the definition of a geodesic, we can erect a tetrad $\{e_i\}$ such that
\begin{equation} ds=c\frac{dt}{d\tau}d\tau-\frac{dx}{ds}dx-\frac{dy}{ds}dy-\frac{dz}{ds}dz\label{ds}\end{equation}
along a geodesic. Sometimes, we will prefer to write this in the form
\begin{equation} cds=Hdt-p_1dx^1-p_2dx^2-p_3dx^3
\label{cds}\end{equation}
where $p_o$ is denoted by $-H$, $c$ is the speed of light in a vacuum and $p_i=c({dx^i}/{ds})$. Note that since 
$ds$ is an exact differential $s$ has the form
of a Hamilton-Jacobi function. In particular, if we change the parametrization from $\tau$ to $\tau^*$ then 
equation (\ref{cds}) can be rewritten as
\begin{equation}
cds^*=H^*dt-p^*_1dx^1-p^*_2dx^2-p^*_3dx^3\label{cds^*}~.
\end{equation}
Equations (\ref{cds}) and (\ref{cds^*}) can be interpreted in two ways. From one perspective they represent two different geodesics parameterized with respect to a common time parameter; from another they can be seen as representing the same curve with respect to different tetrads. The important thing is that in either interpretation we can pass from one representation to the other by means of the expression $ds=\frac{ds}{ds^*}{ds^*}.$
For our purposes, it is sufficient to erect a tetrad in the rest frame of the particle and define a time parameter $\tau$ with respect to a standard clock
in this frame.

In the case of a piecewise geodesic $ds=\sum ds_i$ where each $ds_i$ defines a line segment on a geodesic 
with respect to the laboratory tetrad, each $s_i$ can be parameterized in terms of $s_1$, say, which in 
turn can be parameterized by $\tau$. In other words, each component $s_i$ is a Hamilton-Jacobi function
which can be synchronized with the laboratory frame by means of the expression
$$
\frac{ds_i}{d\tau}=\frac{ds_i}{ds_1}\frac{ds_1}{d\tau} .
$$ 
Note $\frac{ds_i}{d\tau}$ does not have to be a constant. In particular, if $\tau$ is chosen as a standard 
time parameter then all geodesics can be parameterized with respect to the same parameter by means of this 
relationship. This is equivalent to defining a universal time parameter \cite{ohara}, first introduced by 
Stuekelberg in 1941, \cite{stk}, and further developed by Horwitz et al.\ \cite{hor}.

With a universal laboratory time established, we now derive two other equations from the metric which will be useful later on for describing specific
realizations of Brownian processes. Equation (\ref{ds^2}) expressed in tetrad notation becomes
$$
ds^2=c^2dt^2-dx^2-dy^2-dz^2.
\label{4}
$$
Letting $\dot{s}={ds}/{d\tau}$ and $\ddot{s}={d^2s}/{d\tau^2}$ this can be rewritten with respect to the 
time parameter $\tau$ as
\begin{equation}
\dot{s}ds=c^2\dot{t}dt-\dot{x}dx-\dot{y}dy-\dot{z}dz~.
\label{sdotds}
\end{equation}
In the case where the standard time $\tau$ is a non-affine\footnote{For an affine parameter there is no acceleration which means $\ddot{s}=0$.}
parameter of the local time $t$, the second derivative does not vanish and we can write
\begin{equation}
\dot{s}\ddot{s}=c^2\dot{t}\ddot{t}-\dot{x}\ddot{x}-\dot{y}\ddot{y}-\dot{z}\ddot{z}
\label{6}
\end{equation}
or equivalently
\begin{equation}
\ddot{s}ds=c^2\ddot{t}dt-\ddot{x}dx-\ddot{y}dy-\ddot{z}dz
\label{ddots}
\end{equation}
or
\begin{equation}
\ddot{s}ds=c^2\ddot{t}dt-\dot{x}d\dot{x}-\dot{y}d\dot{y}-\dot{z}d\dot{z},
\label{ddots1}
\end{equation}
where equation (8) follows from (7) by observing that for any function $f$
$$
df \frac{d^2 f}{d \tau^2} = d f \frac{d}{d \tau} \frac{d f}{d \tau} = \dot f d \dot{f}.
$$
Note that since $s(\tau)$ is a scalar function, $\ddot{s}ds=d\dot{s}^2$ is both Lorentz invariant by
construction and also an exact differential with respect to $\tau$.

\section{Brownian Motion}
To investigate a possible notion of relativistic Brownian motion, let us return to equation
(\ref{sdotds})
$$
\dot{s}ds=c^2\dot{t}dt-\dot{x}dx-\dot{y}dy-\dot{z}dz ~,
\label{ddnew}
$$
defined with respect to a universal time parameter $\tau$
at any point along a curve. For what follows, we will restrict the motion to piecewise geodesic 
curves with $Var(\tau_i-\tau_{i-1})>0$;
in other words, to those curves for which the tangent is defined at every point along a (smooth) piece
of the trajectory geodesic in $\cal M$ in between two instantaneous random events occurring at 
discrete times $\zt_{i-1}$ and $\zt_i$, which deviate it from its free motion\footnote{because 
of collisions with other particles or with various kinds of obstacles
\cite{MR94,LNRM,CCFV,JR06}.} such that
$$
s(\zt)=\sum^{n-1}_{i=1}\int_{s_{\tau_{i-1}}}^{s_{\tau_i}} ds_i + \int^{s_{\tau}}_{s_{\tau_{n-1}}}ds_{n}~.
$$
To put this into the context of probability theory, we now need to define random variables on the sample space
$\Omega=\{s\}$ of timelike geodesics on $\cal M$.
Let $S$, $\dot{S}$ and $\ddot{S}$ be three random variables on $\Omega$ defined by
$$
S(s)=s,\qquad \dot{S}(s)=\frac{ds}{d\tau}\qquad \textrm{and}\qquad \ddot{S}(s)=\frac{d^2s}{d\tau^2}\qquad
\mbox{with }~~
\tau \in [0,\infty)~.
$$
On occasion, we will also write $S=(T,X,Y,Z)$ or $S=X^a,\ a\in \{0,1,2,3\}$ where each $X^a$ is a projected cosine
of $S$ with respect to a tetrad.
In addition, we can define a random differential $dS$ by $\ dS(s(\tau))=ds$ and a random variable
\begin{equation}
\dot{S}dS=c^2\dot{T}dT-\dot{X}dX-\dot{Y}dY-\dot{Z}dZ = X^adX_a~,  \quad \mbox{with }~~
\tau \in [0,\infty)~,
\end{equation}
which for each event $s$ can be associated with the metric (\ref{sdotds}).
Also, in the event that motion along the geodesic is not at a constant speed with respect 
to the standard clock, a random variable associated with acceleration
can be identified with equation (7) for each realization of the motion. It is given by
\begin{equation}
\ddot{S}dS=c^2\ddot{T}dT-\ddot{X}dX-\ddot{Y}dY-\ddot{Z}dZ= \ddot{X}^adX_a
\label{ddots2}
\end{equation}

We now introduce a set of stopping time $\{\tau_i\}$ and a set of random variables $S_i$, $\dot{S}_i$, 
$\ddot{S}_i$, indexed by the stopping times, and
defined over the sample space of timelike geodesics $\{s(\tau)\}$ such that for all $\tau \in [\tau_{i-1},\tau_i)$
$$
S_i(s)=s(\tau_i-\tau_{i-1}),\ \dot{S}_i(s)=\frac{ds}{d\tau}(\tau_i)\ \textrm{and}\ \ddot{S}_i(s)=\frac{d^2 s}{d\tau^2}(\tau_i).
$$

In particular, in the case of a piecewise smooth curve $\sigma(s)=\bigcup \sigma_i(s)$ where $\sigma_i(s)$ is a line segment lying on a geodesic with differential $ds_i$, indexed by the stopping times, the overall length of a trajectory of a particle undergoing Brownian motion during the time interval $[0, \tau)$
where $\tau \in [\tau_{n-1},\tau_{n})$, can be re-expressed as:
\begin{equation}
s(\zt)=\sum^{n-1}_{i=1}\int_{S_{\tau_{i-1}}}^{S_{\tau_i}} ds_i + \int^{S_{\tau}}_{S_{\tau_{n-1}}}ds_{n}
=\sum^{n-1}_{i=1}\int_{S_{\tau_{i-1}}}^{S_{\tau_i}} dS_i(\sigma_i(s)) + \int^{S_{\tau}}_{S_{\tau_{n-1}}}dS_{n}(\sigma_n(s)).
\end{equation}
\newline

It follows that for each realization of the processes $\{\tau_i\}_{i=1}^\infty$ defined with respect to the standard clock in the rest frame of the laboratory, $\{s_i\}_{i=1}^\infty$, $\{\dot{s}_i\}_{i=1}^\infty$ and $\{\ddot{s}_i\}_{i=1}^\infty$ define
Markov random walks with respect to the (well ordered) index set of stopping times $\tau_i$. In all we have at 
least five different random variables indexed by the stopping times: the difference between stopping times 
($\chi_i=\tau_i-\tau_{i-1})$, the length traveled between stopping times ($S_i$),
the initial speed $|V^a_{i}|$) at each stopping time, the initial acceleration magnitude ($|F^a_{i}|$) at each 
stopping time and the unit
velocity (direction of the trajectory) $\left(\frac{V^a}{|V^a|}e_a\right)_i$ with respect to the laboratory 
inertial frame at each stopping time.\footnote{In the geodesic parts of the motion the direction of velocity
equals the direction of acceleration.}

Suppose now that the heat bath is characterized by a unique constant parameter, the temperature $\Theta$,
as in the classical case. Then all five families of random variables can be considered to be stationary 
independent processes and the specific form of the resulting Markov processes will depend on the distributions of these 
random variables. 
Also, in this case $\{S_i\}_{i=1}^\infty$, $\{\dot{S}_i\}_{i=1}^\infty$ and $\{\ddot{S}_i\}_{i=1}^\infty$ will have 
infinitely divisible distributions which can be associated with Levy processes. 

The details of the relevant statistics depend on the properties of the environment in which the process takes 
place, commonly known as the heat bath. As usual in Brownian motion theory, we assume that the only relevant 
parameter of the heat bath is its temperature $\Theta$. Consequently, for constant temperature $\Theta$, 
$S=\sum_{i=1}^n S_i$, $\dot{S}=\sum_{i=1}^n \dot{S}_i$ and $\ddot{S}=\sum_{i=1}^n \ddot{S}_i$ converge in 
distribution to stable processes. Indeed, this  $\Theta$-dependence will influence
the time increments $\chi$, and the measure of a universal unit of time.
In what follows, we assume (as suggested by molecular dynamics studies such as \cite{ARM,GM11})that the 
underlying thermal bath state is characterized by the isotropy of test particle trajectories associated 
with the independent and identically distributed (with $\Theta$-dependent distribution) {\em time-like}
increments of stopping times $\chi_i$.

With notation clarified, we now construct a statistical model for relativistic Brownian motion by examining 
two cases.\footnote{As it turns out, in any inertial frame an infinite number
of Markov processes (one for each $k$) can be defined along a geodesic by
$$
\biggl\{\frac{d^k S_i}{d\tau^{k}}\left|\frac{d^k S_i}{d\tau^k}- \frac{d^k S_{i-1}}{d\tau^k}\ \right.
\textrm{are stationary and independent},\ i \in  \zN \biggr\},
$$
provided the derivatives are not everywhere equal to zero.}

\subsection{Relativistic Gaussian Distribution:} Consider the family of stopping times $\{\tau_i:i\in \mathbb{N}\}$, assuming that the set of
random variables $\{\chi_i=\tau_i-\tau_{i-1}\}$ are independent and identically distributed. The fundamental
metric is
$$
ds^2=c^2dt^2-dx^2-dy^2-dz^2
$$
which can be differentiated with respect to the parameter $\tau$ yielding Eq.(\ref{sdotds}).
This metric can be associated with the quaternions by means of the inner product
$$
\bigl<q_1,q_2\bigr>\equiv \frac14(q_1q_2+q^*_1q^*_2+q_2q_1+q^*_2q^*_1)=x_oy_o-x_1y_1-x_2y_2-x_3y_3
$$
which is equivalent to the Minkowski inner product, where $q_1$ and $q_2$ are quaternions
and $q^*_1,q^*_2$ their conjugates. Moreover, since each quaternion is also
a four vector on Minkowski space, we can define a Lorentz transformation $q^{\prime}=Aq$ such that
\begin{eqnarray*} \bigl<q^{\prime}_1,q^{\prime}_2\bigr>&=&x^{\prime}_oy^{\prime}_o-x^{\prime}_1y^{\prime}_1-x^{\prime}_2y^{\prime}_2-x^{\prime}_3y^{\prime}_3\\
&=&x_oy_o-x_1y_1-x_2y_2-x_3y_3\qquad \textrm{by Lorentz transformation}\\
&=&\bigl<q_1,q_2\bigr>.
\end{eqnarray*}

Indeed, written in this way the quaternions together with this inner product can be identified with
Minkowski space. This now enables us to associate a random variable $S_i=S(\chi_i)$
and random vectors $S=T \zII +Xj+Yk+Zl$ with each stopping time $\tau_i$ such that
$$
\{(T,X,Y,Z)\in Q| d\mathbf{S}^2=c^2dT^2-dX^2-dY^2-dZ^2\}
$$
where $\zII, j, k, l$ are the standard basis for $Q$, the set of quaternions.
If between stopping times a particle is moving with velocity
$v^a_j(\tau)=(c\dot{t}, \dot{x},\dot{y}, \dot{z})\in Q$, along a geodesic then the distance traveled componentwise between stopping times $(\tau_{i-1},\tau_i)$ is given by
\begin{equation}
x(\chi_i)^a=\int^{\tau_i}_{\tau_{i-1}}v^ad\tau > 0,\label{dis}
\end{equation}
which, because of the randomly assigned initial velocities at each $\tau_i$, can be rewritten as a stochastic random vector (quaternion)
\begin {equation}
X^a_i=\int^{\tau_i}_{\tau_{i-1}}V^a_{i-1}(\tau)d\tau,
\label{14}
\end{equation}
where $X^a_i=X^a(\chi_i)$.
It follows by isotropy that for each $a\in \{1,2,3\}$, $\zE(X^a_i)=\zE(X^a_ie_a)=0$ and it also follows from isotropy and (\ref{dis}) that
$Var(X^1_i)=Var(X^2_i)=Var(X^3_i)>0$ exists.
In particular, if $V^a_{i-1}(s)$ is constant along $s$ then
\begin {equation}
X^a_i=V^a_{i-1}\chi_i,
\label{16}
\end{equation}
and
\begin{equation}
Var(X^a_i)=Var(v_{i-1})Var(\tau_i-\tau_{i-1})=\sigma^2(v)\sigma^2(\chi)
\end{equation}
exists.

However, it is not immediately clear how we should interpret the time variable. First note that $\chi_i=\tau_i-\tau_{i-1}=-(\tau_{i-1}-\tau_i)$. Seen from the perspective of the event $\tau_{i-1}$, $\chi_i$ equals the time between the $i-1$ kick and the future kick at $\tau_i$. Therefore, $\zE(|\chi_i|)=\mu_t$ is the mean future time to the next event. However, seen from the perspective of $\tau_i$, $\chi_i$ equals the time between the $i$ kick and the previous kick at $\tau_{i-1}$. Therefore $\zE(-|\chi_i|)=-\mu_t$ represents the mean past time to the previous kick. Consequently it is meaningful to ask what is the probability that the next kick will take place within $t$ seconds and it is also meaningful to ask what is the probability that the previous kick took place within the past $t$ seconds. Taking the two statements together, we can maintain the arrow of time going from past to future, while meaningfully extending the range of $\chi$ to $(-\infty, \infty)$.  For this extended range of $\chi$, it again follows by symmetry that $\zE(\chi_i)=0$.

In particular for the random variable vector
\begin{equation}
S^*_n=\sum^n_{i=1} X^a_i = \sum^n_{i=1}S_i
\end{equation}
composed of a sum of i.i.d.\ variables $X^a_{\tau_k}-X^a_{\tau_{k-1}}$ we can associate a probability
measure over the quaternions given by
$$
dP(n, x^a)=P(S^*_n \in (ct,ct+cdt)\times (jx, jx+jdx))\times (ky,ky+kdy)\times (lz,lz+ldz)).
$$
By the central limit theorem $S^*_n$ is asymptotic for large $n$ to the (quaternion) Gaussian
distribution in Minkowski space, with variance growing linearly with $n$, given by
\begin{equation}
f_s(t\zII+jx+ky+lz;n) \sim \frac{1}{4\pi^2 n^2 \sigma_{\chi}\sigma_{X}\sigma_{Y}\sigma_{Z}}
\exp\left\{-\frac{1}{2n}\left(\frac{c^2t^2}{\sigma^2_{\chi}}-
\frac{x^2}{\sigma^2_X}-\frac{y^2}{\sigma^2_Y}-\frac{z^2}{\sigma^2_Z}\right)\right\} ~,
\label{Qgauss}
\end{equation}
which holds for $c^2t^2-x^2-y^2-z^2 = O(n)$ or less.

As noted above, for the quaternion 
$$
q = -\frac{1}{\sqrt{2n}}
\biggl(\zII\frac{ct}{\sigma_{\chi}}-j\frac{x}{\sigma_X}-k\frac{y}{\sigma_Y}-l\frac{z}{\sigma_Z}\biggr)
$$
$s^2=\left<q,q\right>$ is invariant. 
This can also be expressed in tensor notation by $x^{i^{\prime}}=a^i_jx^j$, where $A=a^i_j$ is a
Lorentz transformation such that:
$$
s^2=\Sigma_{i^{\prime}j^{\prime}}x^{i^{\prime}}x^{j^{\prime}}=
a^{i^\prime}_ia^{j^{\prime}}_j\Sigma_{ij}a^i_{i^\prime}x^i a^j_{j^{\prime}}x^j=\Sigma_{ij}x^ix^j~.
$$
It can also be written more succinctly as
$$s^2=(q^\prime)^{\dag}\Sigma^{\prime}q^{\prime}=(Aq)^{\dag}A\Sigma A^{\dag}Aq=q^{\dag}\Sigma q,$$
where $\Sigma$ is the covariant matrix of the vector in Minkowski space.

\subsection{Diffusion Equation}
Before writing down the diffusion equation for the above process,
it is important to note that there are two time parameters expressed in Eq.(\ref{Qgauss})
and both are
independent of each other. The continuous $\chi$ parameter is associated with the {\em systematic}
geodesic motion between two stopping times, the other with the {\em randomly occurring discrete} set of stopping times $\{\tau_i \}$. The first is directly related to the proper times between stopping times and is used to index the
deterministic part of the process, while the index $n$ associated with the stopping times is used to index the relativistic space-time stochastic processes. Once these distinctions are kept in mind the usual paradox associated with interpreting
random walks, whereby there is a finite probability that the Brownian particle may be in
superluminous regions, no longer exists. Indeed, Eq.(\ref{Qgauss}) was constructed with the assumption that all deterministic motion
is along timelike geodesics. Consequently the motion can never be in a superluminous region.

With this distinction made, one observes that a relativistic Brownian particle obeys the following kind 
of ``diffusion'' equation:
\be
\frac{\sigma^2_{\chi}}{2c^2}\frac{\partial^2 F}{\partial t^2}-\frac{\sigma^2_X}{2}\frac {\partial^2 F}{\partial x^2}-\frac{\sigma^2_Y}{2}
\frac{\partial^2 F}{\partial y^2}-\frac{\sigma^2_Z}{2}\frac{\partial^2 F}{\partial z^2}-
\frac{\partial F}{\partial n}=0.\
\label{Teleg}
\ee
This formally resembles the telegraph equation on Minkowski space as found for example in Ref.\cite{Keller}.
However, it is radically different in that
the first and second order time derivatives of (\ref{Teleg})
are taken with respect to different time variables. Moreover, there is no bound on the occurrences of stopping times. The time and the distance traveled between
stopping times may be arbitrarily large, as long as they are consistent
with speeds smaller than $c$. This generalizes the approach of e.g.\ Ref.\cite{Keller},
in which the stopping times are deterministically generated and equal $n$ at the $n$-th collision, 
for a one dimensional random walk that increases or decreases incrementally by one unit during each 
time interval. 
In other words, the case discussed by Keller is a special instance of our time-like condition, in which
our two time variables become a unique variable, leading to the standard telegraph equation for 
the propagation of the particle.

\vskip 5pt
\subsection{\bf{Maxwell-J\"uttner Distribution:}} If instead of Eq.(\ref{sdotds}) 
we begin with the metric Eq.(\ref{ddots1}):
$$
\ddot{s}ds=c^2\ddot{t}dt-\dot{x}d\dot{x}-\dot{y}d\dot{y}-\dot{z}d\dot{z}
$$
we can now repeat the above theory by replacing Eq.(\ref{14}) with the equation
\begin{equation}
\dot{X}^a_i=V(\tau_i)-V{(\tau^-_i)}+ \int^{\tau^-_i}_{\tau_{i-1}}F^a_{i-1} d\tau
\end{equation}
where $V(\tau_i)-V{(\tau^-_i)}$ defines the random discontinuity in velocities that occur at each stopping time and $F^a_{i-1}$ represents the four-acceleration along the geodesic segment from time $\zt_{i-1}$ to
time $\zt_i$. For example $F^a_{i-1}$ could be the acceleration of a Brownian particle performing a simple harmonic motion betweens kicks.
The above can now be repeated to generate a distribution which formally looks
like the Maxwell-J\"uttner distribution of special relativisitc gases:
\begin{equation}
f_v(\zII\dot{t}+j\dot{x}+k\dot{y}+l\dot{z}) \sim
\frac{1}{(2\pi n)^2|\Sigma|}
\exp\biggl(-\frac{1}{2n}(\frac{c^2\dot{t}^2}{\sigma^2_{\dot{T}}}-\frac{\dot{x}^2}{\sigma^2_{\dot{X}}}-\frac{\dot{y}^2}{\sigma^2_{\dot{Y}}}-
\frac{\dot{z}^2}{\sigma^2_{\dot{Z}}})\biggr).
\label{fv}
\end{equation}
but because of the two time parameters $t$ and $n$, it is radically different from the Maxwell-J\"uttner
distribution, and it applies to accelerating particles. In any event, $f_v$
is clearly covariant since $|\Sigma|=\sigma_{\dot{T}}\sigma_{\dot{X}}\sigma_{\dot{Y}}\sigma_{\dot{Z}}$ is invariant.

An interesting special case occurs when there are no accelerations along the geodesic 
and velocity is a constant. In this case, equation (19) reduces to
\begin{equation}
\Delta \dot{X}^a_i=V(\tau_i)-V(\tau^-_i)
\end{equation}
and the random variable $\dot{S}^*_v\equiv \sum^n_{i=1}V^a_i$ defines a sum of four independent and stationary random vectors over the quaternions and converges to the distribution given in equation (20). Clearly the same argument can be extended to any random variable such as acceleration that undergoes a change from $\tau^-_i$ and $\tau_i$.

The above derivation suggests that at the core of the molecular chaos hypothesis in relativity are the stationary 
independent random velocities (and accelerations) produced at stationary and independent random stopping 
times. This would need to be investigated more in depth,
since it departs sensibly from the standard wisdom on the Brownian motion.

\section{Concluding remarks}

\begin{enumerate}
\item It is important to emphasize that in our calculation in order to have a probability distribution, integration is carried out over the quaternions
and not $\zR^4$. For example
\begin{eqnarray*}P(T\in(t_1,t_2), X\in(a,b))&=&\frac{1}{2\pi n}\int^{t_1}_{t_0}\int^{ib}_{ia}e^{-\frac{t^2}{2n\sigma^2_T}+\frac{x^2}{2n\sigma^2_X}}dt d(ix)\\
&=&\frac{1}{2\pi n}\int^{t_1}_{t_0}\int^{b}_{a}e^{-\frac{t^2}{2n\sigma^2_T}-\frac{x^2}{2n\sigma^2_X}}dt dx
\end{eqnarray*}
This is equivalent to have constructed a Markov
chain on a Minkowski manifold subjected to the conditions of special relativity.
\item In the above formulation we have extended this to pseudo-Riemannian manifolds by defining the geodesic with respect to a fixed tetrad, which means that locally in can be identified with Minkowski space.
\item In both the Brownian motion and the Maxwell J\"uttner distribution, the above theory can be extended to non-geodesic smooth curves $W(s)$ parameterized by the geodesic curve length $s$ by means of the relation $dW=\frac{dW}{ds}ds$.
\item Clearly $f_s(\zII t+jx+ky+lz)$ is invariant under the (homogeneous) Lorentz group, 
while $f_v(\zII\dot{t}+j\dot{x}+k\dot{y}+l\dot{z})$ is invariant under Poincare group.
\item The distribution $f_s$ and $f_v$ can be factorized into separate time and space component products. For example
$$
f(\zII t+jx+ky+lz)=f_{n}(t)f_{\bf{X}}(jx+ky+lz),$$ where $f_{\bf{X}}(jx+ky+lz)$ is Gaussian.

\end{enumerate}

The above, leads to a characterization of the equilibrium state in the general relativity framework
which, along the lines of classical statistical mechanics, requires almost no information about the
details of the interactions among the objects of interest. Nevertheless, these details are important
and implicit in the above formulation.
In the first place they are the underlying cause of the unpredictable motions associated with the
random interaction times and are
constitutive of the phenomenon. In the second place, the statistics of these
interactions depend on the temperature $\Theta$. Knowledge of these details is implicitly contained in the
relation between the stopping times and the laboratory frame time, and should be made explicit to
fully characterize the relativistic Brownian motion.

\section*{Acknowledgments}
P.O.H.\ is grateful for the hospitality at Politecnico di Torino where part of this research has
been carried over.
L.R.\ gratefully acknowledges the warm hospitality provided by NEIU,
as well as financial support from the European Research Council under the European Community's
Seventh Framework Programme (FP7/2007-2013)/ERC Grant agreement No. 202680.


\begin{thebibliography}{99}
\bibitem{AE05} A.\ Einstein, {\it On the movement of small particles suspended in a stationary liquid
demanded by the molecular-kinetic theory of heat}, Ann.\ Phys.\ {\bf 17}, 549 (1905).
{\it On the theory of the brownian movement}, Ann.\ Phys.\ {\bf 19}, 371 (1906).
\bibitem{Dup} B.\ Duplantier, {\it Brownian Motion, ``Diverse and Undulating''},
Progr.\ Math.\ Phys.\ {\bf 47}, 201 (2006)
\bibitem{BPRV08} U.\ Marini Bettolo Marconi, A.\ Puglisi, L.\ Rondoni, A.\ Vulpiani,
{\it Fluctuation–dissipation: Response theory in statistical physics}, Phys.\ Rep.\ {\bf 461}, 111 (2008)
\bibitem{dth07} J.\ Dunkel, P. Talkner and P.\ H\"anggi, {Relativistic diffusion processes and random walk models}, Phys.\ Rev.D{\bf 75}, 043001, (2007)
\bibitem{DH09} J.\ Dunkel, P.\ H\"anggi, {\it Relativistic Brownian motion}, Phys.\ Rep.\ {\bf 471}, 1 (2009)
\bibitem{CCFV} F.\ Cecconi, D.\ del-Castillo-Negrete, M.\ Falcioni and A.\ Vulpiani,
{\it The origin of diffusion: the case of non-chaotic systems}, Physica D {\bf 180}, 129 (2003)
\bibitem{MR94}G.P.\ Morriss, L.\ Rondoni, {\it Periodic Orbit Expansions for the Lorentz Gas}, J.\ Stat.\
Phys.\ {\bf 75}, 553 (1994)
\bibitem{LNRM} J.\ Lloyd, M.\ Niemeyer, L.\ Rondoni, G.P.\ Morriss, {\it The Nonequilibrium Lorentz Gas}, CHAOS{\bf 5}, 536  (1995)
\bibitem{JR06} O.G.\ Jepps, L.\ Rondoni,{\it Thermodynamics and complexity of simple transport phenomena.} In: J.\ Phys.\ A {\bf 39}, 1311  (2006)
\bibitem{ARM} A.\ Aliano, L.\ Rondoni, G.P.\ Morriss, {\it Maxwell-J\"uttner distributions in
relativistic molecular dynamics}, Eur.\ Phys.\ J.\ B {\bf 50}, 361 (2006)
\bibitem{GM11} M.\ Ghodrat, A.\ Montakhab, {\it Molecular dynamics simulation of a relativistic gas:Thermostatistical properties}, Comp.\ Phys.\ Comm.\ {\bf 182}, 1909 (2011)
\bibitem{hakim68} R.\ Hakim, {\em Relativistic Stochastic Processes}, J.\ Math.\ Phys.\ {\bf 9}, 1805 (1968)
\bibitem{Larral} J.\ Almaguer, H.\ Larralde, {\it A relativistically covariant random walk}, 
J.\ Stat.\ Mech.\ P08019  (2007)
\bibitem{GGbook} G.\ Gallavotti, {\em Statistical Mechanics: A Short Treatise}, Springer, Berlin (2000)
\bibitem{CFLV} P.\ Castiglione, M.\ Falcioni, A.\ Lesne, A.\ Vulpiani, {\it Chaos and Coarse Graining
in Statistical Mechanics}, Cambridge University Press, Cambridge (2008)
\bibitem{DGRel} S.R.\ de Groot, W.A.\ van Leeuwen, Ch.G.\ van Weert, {\it Relativistic Kinetic Theory}
North-Holland, Amsterdam (1980)
\bibitem{CCRel} C.\ Cercignani, G.M.\ Kremer, {\it The Relativistic Boltzmann Equation: Theory and
Application}, Birkh\"auser, Basel (2000)
\bibitem{ohara} P.\ O'Hara,{\it Constants of the Motion, universal time, and the Hamilton-Jacobi Function in General Relativity} to appear Jour. Phys: Conf. Series (2013).
\bibitem{stk}Steuckelberg,E.C.G. Helv. Phys. Acta  14 (1941) 322.
\bibitem{hor} Horwitz, L.P. and C. Piron, Helv. Phys. Acta 46 (1973) 316
\bibitem{Keller} J.B.\ Keller, {\em Diffusion at finite speed and random walks}, PNAS {\bf 101}, 1120 (2004)



\end{thebibliography}
\end{document}